\documentclass{aastex62}

\graphicspath{{./}{figures/}}

\received{\today}

\begin{document}

\title{Intensity Interferometry observations of the H$\alpha$ envelope of $\gamma$\,Cas with M\'{e}O and a portable telescope}

\correspondingauthor{Nolan Matthews}
\email{nolankmatthews@gmail.com}

\author{Nolan Matthews}
\affil{Universit\'{e} C\^{o}te d'Azur, CNRS, Institut de Physique de France, France}

\author{Jean-Pierre Rivet}
\affiliation{Universit\'{e} C\^{o}te d'Azur, Observatoire de la C\^{o}te d'Azur, CNRS, Laboratoire Lagrange, France}

\author{David Vernet}
\affiliation{Universit\'{e} C\^{o}te d'Azur, Observatoire de la C\^{o}te d'Azur, CNRS, UMS Galil\'{e}e, France}

\author{Mathilde Hugbart}
\affil{Universit\'{e} C\^{o}te d'Azur, CNRS, Institut de Physique de France, France}

\author{Guillaume Labeyrie}
\affil{Universit\'{e} C\^{o}te d'Azur, CNRS, Institut de Physique de France, France}

\author{Robin Kaiser}
\affil{Universit\'{e} C\^{o}te d'Azur, CNRS, Institut de Physique de France, France}

\author{Julien Chab\'{e}}
\affil{Universit\'{e} C\^{o}te d'Azur, Observatoire de la C\^{o}te d'Azur, CNRS, Laboratoire G\'{e}oazur, France}

\author{Cl\'{e}ment Courde}
\affil{Universit\'{e} C\^{o}te d'Azur, Observatoire de la C\^{o}te d'Azur, CNRS, Laboratoire G\'{e}oazur, France}

\author{Olivier Lai}
\affil{Universit\'{e} C\^{o}te d'Azur, Observatoire de la C\^{o}te d'Azur, CNRS, Laboratoire Lagrange, France}

\author{Farrokh Vakili}
\affil{Universit\'{e} C\^{o}te d'Azur, Observatoire de la C\^{o}te d'Azur, CNRS, Laboratoire Lagrange, France}

\author{Olivier Garde}
\affil{2SPOT (Southern Spectrocopic Project Observatory Team)}
\affil{Observatoire de la Tourbi\`{e}re - 38690 Châbons - France}

\author{William Guerin}
\affil{Universit\'{e} C\^{o}te d'Azur, CNRS, Institut de Physique de France, France}



\begin{abstract}

We report on observations of the extended environment of the bright Be star $\gamma$-Cas performed using intensity interferometry measurements within its H$\alpha$ emission line.
These observations were performed using a modified version of the I2C intensity interferometry instrument installed onto the 1.54\,meter M\'{e}O optical metrology telescope and a portable 1-meter telescope (T1M). 
In order to better constrain the extent of the H$\,\alpha$ envelope, observations were performed for two different positions of the T1M telescope, corresponding to an intermediate and long baselines in which the extended region was partially and fully resolved. 
We find that the observed data are consistent with past interferometric observations of $\gamma$-Cas. 
These observations demonstrate the capability to equip optical telescopes of different optical designs with intensity interferometry capabilities and illustrate the potential to scale a similar system onto many additional telescopes.

\end{abstract}



\section{Introduction} \label{sec:intro}

Recognized as the first stellar object displaying emission line spectra~\citep{Secchi1867}, $\gamma$-Cas is the prototype of the Be stellar class. 
The emission line features originate from radiative processes with up to X-ray energies~\citep{Smith2012} occurring in an extended disc surrounding the star. 
The disc formation is primarily attributed to mass ejection from the central star enabled from a combination of strong radiative pressure, and low effective surface gravity near the equatorial latitudes.
The latter is a consequence of the extremely high rotation rate that is nearly critical, in which the outward centrifugal force is equal to the inward gravitational force.

Due to its bright stellar magnitude and characteristic stellar size, optical interferometry has been extensively used to study the disk emission of $\gamma$-Cas. 
The extended atmosphere of $\gamma$-Cas was first resolved with the I2T interferometer \citep{Thom1986} and subsequently by the GI2T interferometer showing that the $H\alpha$ region could be fit by a disk model and was in Keplerian motion \citep{Mourard1989}. 
Observations by \cite{Quirrenbach1997} with the MkIII interferometer demonstrated that the emission-line region were not compatible with circularly symmetric models and required the assumption of an elongated profile. Density and velocity relationships in the equatorial plane were constrained and accounted for by a radiative wind driven model in \citet{Stee1995}. 
Subsequently, spectro-interferometric measurements of the envelope size were performed across both the H$\alpha$ and H$\beta$ lines, as well as in the near-by continuum emission \citep{Stee1998} leading to a measurement of the disk mass and opening angle \citep{Stee2003}. 
In addition, the Navy Precision Optical Interferometer (NPOI) was used to characterize the disc geometry and further confirmed the oblateness of the disc \citep{Tycner2006}. 
The CHARA interferometric array measured the disk extent in the K' photometric band for the first time, found to be slightly smaller than previous observations in H$\alpha$~\citep{Gies2007}.
Finally, high sensitivity spectro-interferometric measurements with CHARA were performed in the near-infrared, as well across the H$\alpha$ line and near-by continuum suggesting a larger disk size than prior measurements and linking the origin of X-ray emission to a compact binary companion due to the absence of one-armed spiral structures or secondary star~\citep{Stee2012}.

In this work we present the first known intensity interferometry (II) measurements of the extended atmosphere of $\gamma$-Cas using a modified version of our intensity interferometry instrument (I2C) installed onto the 1.54-meter telescope of the M\'{e}O laser ranging facility and a mobile 1-meter telescope (hereafter T1M), both located on the Calern Plateau site of the Observatoire de la C\^{o}te d'Azur. While the I2C instrument shares similarities with past II observations using the telescopes of the Centre P\'{e}dagogique Plan\`{e}te Univers (C2PU) \citep{Guerin2017,Guerin2018,Rivet2020,deAlmeida2022}, there were several modifications required to outfit these telescopes with II capabilities, and to be compatible with each other in an interferometric mode. The experimental setup is thus described in Section~\ref{sec:exp_setup} with additional details also presented in \citet{Matthews2022}. The observations and results are shown in Section~\ref{sec:observations} with an analysis presented in Section~\ref{sec:analysis}. Finally, we discuss the results and present an outlook for future intensity interferometry measurements in Section~\ref{sec:discussion}.

\section{Experimental Setup} \label{sec:exp_setup}

\subsection{Principles of Intensity Interferometry}
\label{sec:principles}

An intensity interferometer correlates the intensity fluctuations of starlight between separated telescopes in order to measure the squared visibility. 
For two telescopes with a projected baseline $r$ between them, the second order coherence function is
\begin{equation}
    g^{(2)} (r,\tau) = \frac{\langle I_1(t) I_2(r,t+\tau)  \rangle}{\langle I_1 \rangle \langle I_2 \rangle}
\end{equation}
where $I_1$ and $I_2$ are the intensities recorded at each of the two telescopes, $\tau$ is the relative time-lag between the signals, and the brackets indicate an average over time $t$. 
The Siegert relation~\citep{Siegert1943,Ferreira2020} relates the second-order coherence function \,$g^{(2)}$\, to the first order coherence function $g^{(1)}$ by
\begin{equation}
    g^{(2)}(r,\tau) = 1 + |g^{(1)}(r,\tau)|^2
\end{equation}
where the first-order coherence function can be separated into spatial and temporal components
\begin{equation}
    g^{(1)}(r,\tau)=V(r) g^{(1)} (\tau)
\end{equation} 
where $V(r)$ is the interferometric visibility of the source, given by the Fourier transform of the source sky brightness distribution. 
For an unresolved point-like source $V(r)=1$, and the resulting second order coherence function will depend only on the temporal component $g^{(1)}(\tau)$ given by the Fourier transform of the measured light spectral density~\citep{Wiener1930,Khintchine1934}. 
For linearly polarized thermal light at zero optical path delay $g^{(1)}(\tau=0) = 1$ where for time-lags much greater than the coherence or correlation time the first order coherence function should be equal to zero such that there is a ``bunching peak" centered about zero optical path delay with an effective temporal width given by the coherence time. 
The peak amplitude at zero time-lag of $g^{(2)}$ thus measures the squared visibility at some projected baseline assuming that the instrumental resolving time is shorter than the light coherence time. The coherence time can be defined by the integral of the squared first-order coherence function~\citep{MandelAndWolf1995},
\begin{equation}
    \label{eqn:coh_time}
    T_c = \int |g^{(1)}(\tau)|^2 d\tau = \int |s(\nu)|^2 d\nu,
\end{equation}
which is equal to the integral of the squared normalized spectral density $s(\nu)$ by Parseval's theorem. For visible light with a bandpass of $\Delta \lambda \sim 1\,$nm the corresponding coherence time is of order 1\,ps, much shorter than what can be achieved with conventional detectors. 
In this case, a measurement averages over many coherence times and reduces the value of the $g^{(2)}$ peak amplitude at $\tau = 0$ by a factor of $\sim T_c / T_d$ where $T_d$ is the effective time-resolution of the detector. 
The amplitude of the $g^{(2)}$ peak therefore measures this loss of contrast times the squared visibility.
The squared visibility can be extracted by dividing the value of $g^{(2)}(r)-1$ peak amplitude measured between telescopes to the $g^{(2)}(r=0)-1$ peak amplitude measured at zero-baseline under the assumption that the profile of the $g^{(2)}(\tau)$ peak is constant. 
In practice, we measure the ratio of the area of the $g^{(2)}(r)-1$ peak to the area of the $g^{(2)}(r=0)-1$ peak for the squared visibility,
\begin{equation}
    \label{eqn:visb2}
    |V(r)|^2 = \frac{\int \left( g^{(2)}(r,\tau)-1 \right) d\tau}{\int \left( g^{(2)}(r=0,\tau)-1 \right) d\tau}.
\end{equation}
The denominator, or equivalently the area of the $g^{(2)}$ peak at zero baseline, corresponds to the coherence time that can be calculated from the measured spectrum as given by Equation~\ref{eqn:coh_time}. 
The equivalence between the coherence time from intensity interferometry and spectral measurements assumes that the spectral resolution is narrower than any spectral lines within the instrumental bandpass.
Since intensity interferometry measurements probe the intrinsic spectrum it is a useful method for characterizing narrow spectral lines present in, for example, studies of light scattering off of atomic clouds~\citep{Dussaux2016}, with potential applications in astrophysics~\citep{Tan2017}.

\subsection{Telescopes}

\begin{figure}
\plottwo{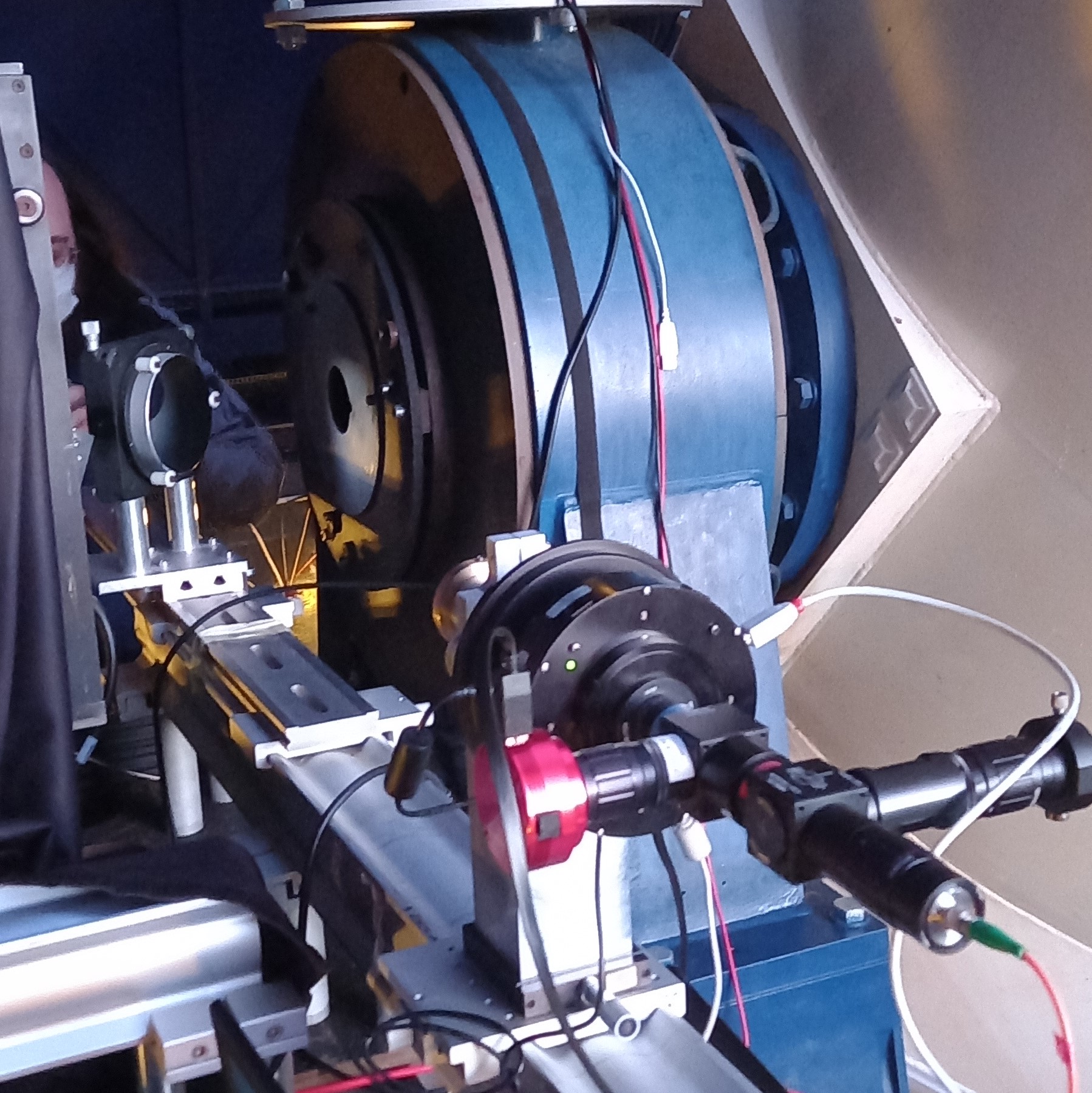}{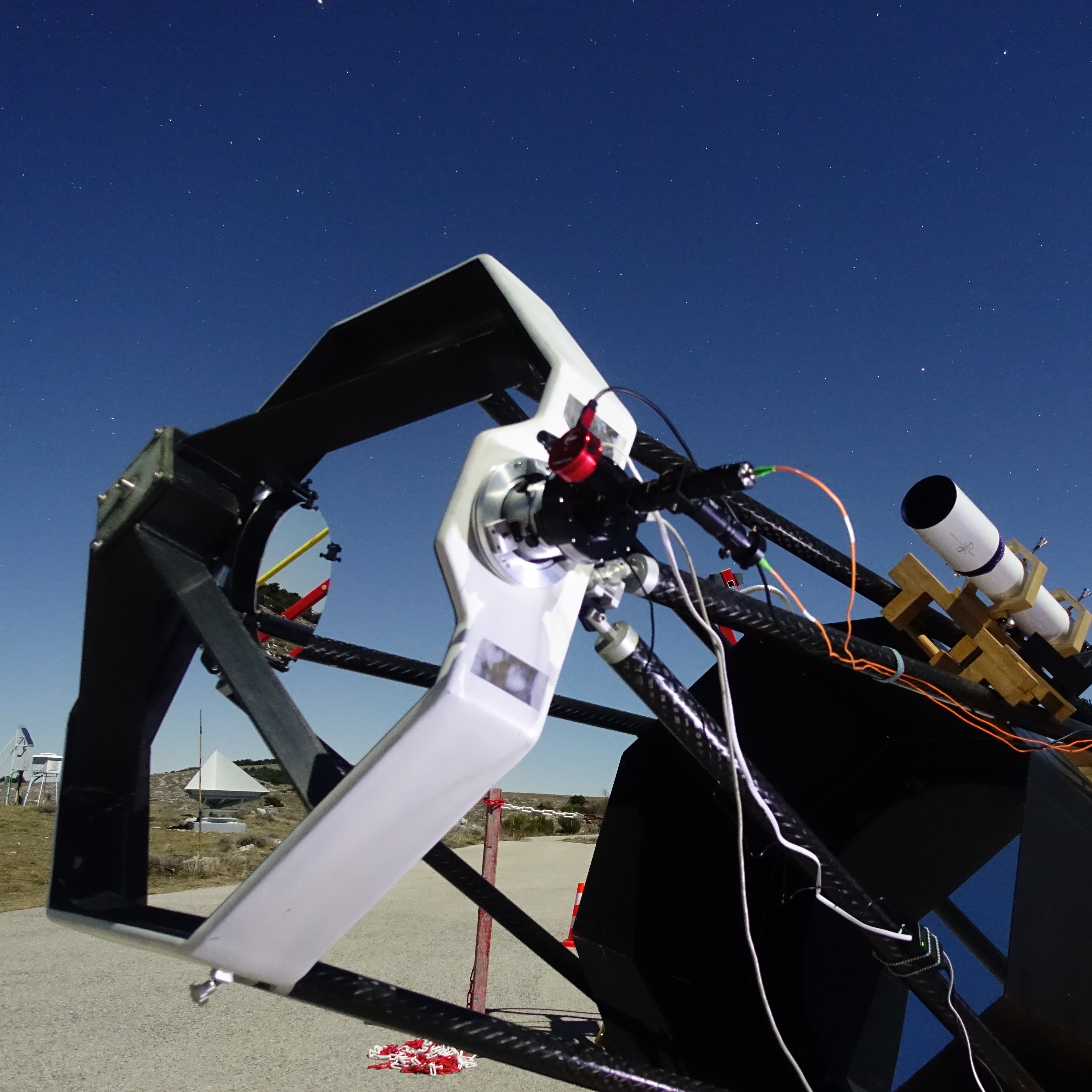}    
\caption{Photographs of the coupling assemblies mounted on the Nasmyth arm of the M\'{e}O telescope (left) and at the Newtonian focus of the T1M portable telescope (right).}
\label{fig:installation_photos}
\end{figure}

The II observations presented in this paper were performed by outfitting two telescopes located on the Calern Plateau site of the Observatoire de la C\^{o}te d'Azur with individual coupling assemblies (CAs).
A CA is mounted near the focus of each telescope, both shown in Figure~\ref{fig:installation_photos}.
The first facility was the 1.54 m diameter M\'{e}O (M\'{e}trologie Optique) telescope primarily used for satellite laser ranging \citep{Bertrand2021}, lunar ranging measurements \citep{Bourgoin2021} and low Earth orbit satellite laser communication \citep{Giggenbach2022}.
The optical design is based upon a Ritchey-Chr\'{e}tien configuration on an altitude-azimuth mount with a primary focal length of 31.0\,m giving an approximate focal ratio of f/20.1.
In typical operation the light is brought to a Coud\'{e} focus, but for II observations the light is redirected to the CA along the Nasmyth arm using a removable 45 degree mirror.
In addition, a f=150\,mm lens is inserted before the CA in order to decrease the effective focal length.

The second facility is the portable 1\,m diameter T1M, a Newtonian telescope on a Dobson-type fully motorized azimuthal mount. 
The portability of the telescope enables configurable baselines to expand the accessible coverage of the uv-plane where the telescope can be disassembled and moved in just a few hours.
The telescope has a primary focal length of 3\,m, and a Barlow lens is included in order to expand the effective length at the input of the CA. 

While both telescopes are azimuthal, there will be a relative field rotation due to the Newtonian versus Nasmyth optical designs. 
However, each CA utilizes polarizing filters that must be aligned with respect to one another, although not with respect to the target as astrophysical polarization effects are not investigated. 
To compensate, the M\'{e}O telescope CA is mounted into a rotation stage that orients the CA such that the polarization axes are aligned. 
The stage is actively controlled throughout the observation where the amount of rotation is determined from the target sky position.

The relative position of both telescopes must be known with a precision less than a few centimeters for optical path delay corrections. For M\'{e}O, the position was previously determined to millimeter accuracy in terrestrial coordinates due to its use in geodetic surveys. 
To determine an absolute position of the mobile T1M telescope, geodetic markers were installed by the National Institute of Geographic and Forest Information at the ground level for several positions and their positions were measured from differential GPS methods.
The T1M was installed above these markers and the offset between the marker and the T1M reference point was estimated. 
The estimated cumulative error on the reference position is $\pm$1.5\,cm in all directions. 

 
\begin{figure}
    \centering
    \includegraphics[width=1.0\linewidth]{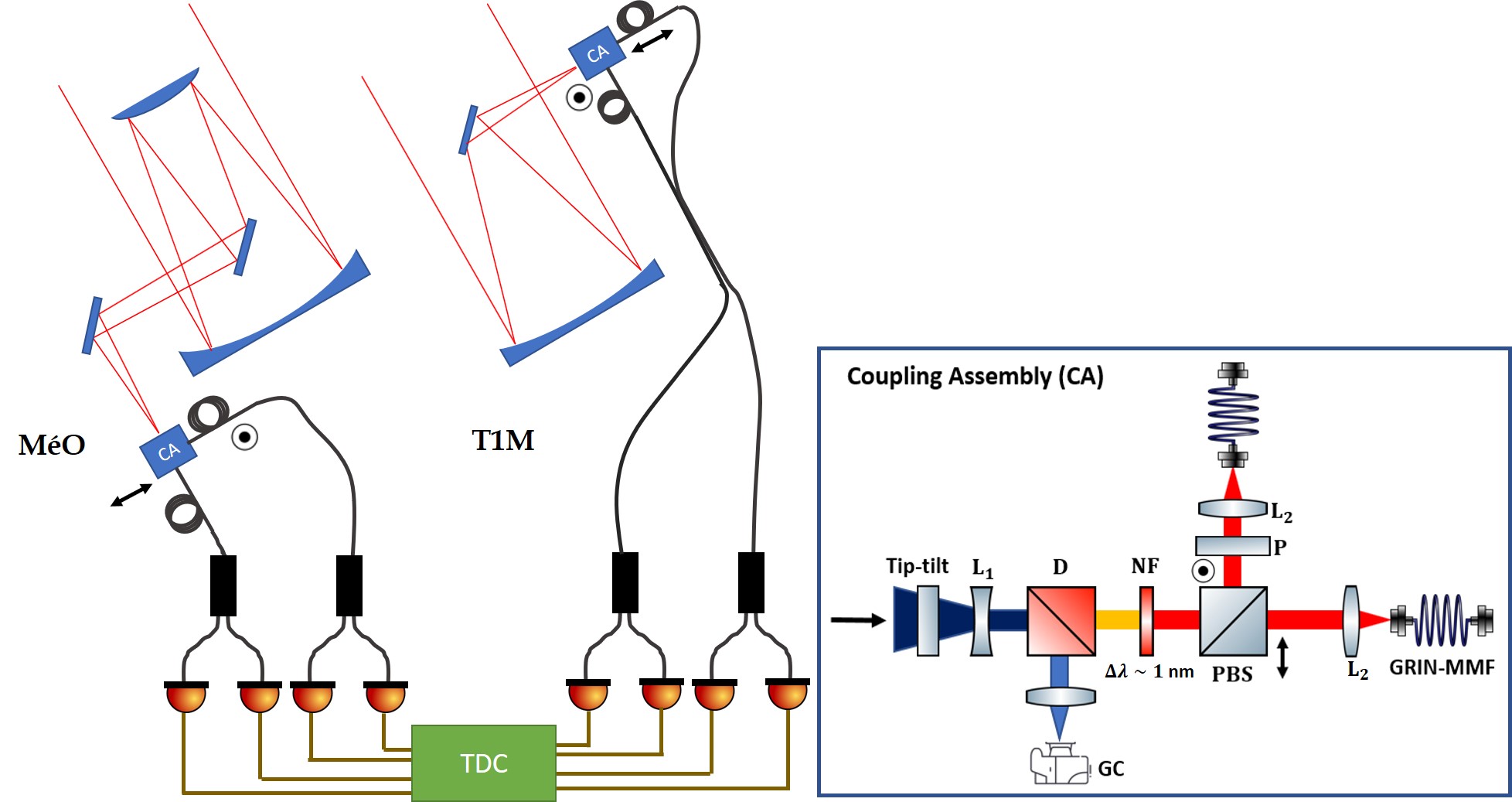}
    \caption{Schematic of the experimental setup. The light collected by both telescopes are brought to individualized coupling assemblies (CA), shown in detail in the right inset. A tip-tilt corrects the beam with respect to transverse displacements. The converging beam is collimated using a diverging lens ($L_1$). A dichroic ($D$) reflects short wavelengths to a guiding camera (GC) used in a closed loop with the tip-tilt. The transmitted light passes through a narrowband filter ($NF$) centered on H$\alpha$. A polarizing beam splitter (PBS) separates the light into orthogonal polarizations where each polarization is injected into a graded-index multimode fiber (GRIN-MMF). A linear polarizer (P) is included on the reflected arm to improve polarization purity. Not shown is the rotation stage used for the M\'{e}O telescope, and additional focal reducers/extenders, which are described in the text. The light for each polarization mode for each telescope is split by a 50/50 fiber beamsplitter, and passed to single photon resolving detectors. The photon arrival times are recorded by the TDC that also produces intensity correlations.}
    \label{fig:tel_schematic}
\end{figure}

\subsection{Instrumental Setup}

The primary function of the CAs (shown in Figure~\ref{fig:tel_schematic}) is to perform spectral/polarization filtering, and fiber injection.
The current version includes an automated tip-tilt device that provides stable fiber injection over several hours without manual intervention \citep{Matthews2022}.
The detector signal output is fed to a time-to-digital converter (TDC) that measures photon arrival times, and produces the correlation between all relevant pairs of detectors.  
Across both telescopes, there are 4 independent measurements of the zero baseline correlation $g^{(2)}(r=0,\tau)$ enabled by using fiber splitters in each polarization mode.
These allow normalization of the spatial correlations across telescopes to measure visibilities. 
Spatial intensity correlations are obtained by calculating the correlation across telescopes for all pairs of detectors in the same polarization mode corresponding to a total of 8 measurements of $g^{(2)}(r,\tau)$. 
During one night, a time delay monitoring system was used on the electronic cables connecting the detectors from the M\'{e}O telescope to the TDC. The drift throughout the whole night was significantly less than the characteristic jitter of the detectors of $\sim 500$\,ps.  

\section{Observations}
\label{sec:observations}

The observations of $\gamma$-Cas were performed between the nights of January 17th, 2022 to January 21st, 2022 in the short baseline configuration, and then from January 24, 2022 to January 27th, 2022 in the longer baseline configuration. 
At the beginning of each night images were recorded in both telescopes of the visual binary system $\gamma$-Ari.
The measured position angle of the binary for both telescopes, and thus of the polarization axes, were found to be always within 5 degrees, corresponding to a loss of visibility of less than 1\%.

\subsection{Temporal Intensity Correlations}

\begin{figure}[b]
\plottwo{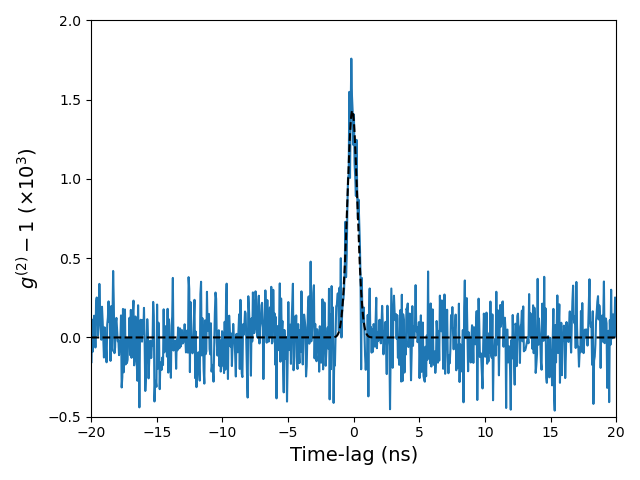}{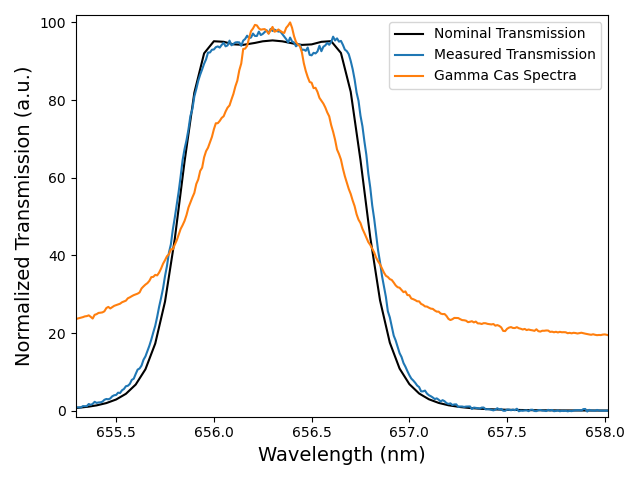}    
\caption{The left plot shows the measured zero-baseline correlation in the blue points with a Gaussian fit shown by the dashed black line. The right image displays the measured spectrum of $\gamma$-Cas, along with the theoretical (black) and measured (blue) spectral transmission of the filter.}
\label{fig:g2_r_0}
\end{figure}

The coherence time obtained from zero baseline intensity correlations were compared to expected values from the spectral throughput.
These temporal intensity correlation functions were computed for each polarization state and for each telescope by summing all individually acquired correlations acquired over the entire observation.
Each of the resulting correlation functions were then shifted by the instrumental delay and then co-added together. 
The resulting correlation function is displayed on the left in Figure~\ref{fig:g2_r_0}. 
The peak is fit by a Gaussian with free parameters for the amplitude and width.
The coherence time, given by the integral of the peak, is extracted via the fit values. 
Before fitting, there is a choice of the time-lag range to fit over and the number of time-lags to bin. 
Here, we fit and show the data over a range of $\pm$\,20\,ns binned into 50\,ps bins.
Under these parameters, we find an amplitude of (1.43$\pm$0.05)$\times 10^{-3}$ and a full-width at half-maximum of 885$\pm$40\,ps corresponding to a measured coherence time of $1.35\pm0.05\,$ps.
Systematics of the fitting process were studied by fitting the data varying the fit range from $\pm$10\,ns to $\pm40$\,ns and additionally the binning size from 10\,ps to 80\,ps. 
Within these parameters we find a maximum difference of 0.015\,ps in the extracted coherence time, notably less than the measurement error,
In the previously quoted coherence time, the correlations from different polarization states and telescopes were co-added and subsequently fit.
This procedure requires that shape of the bunching peak in each correlation, given by the temporal response of the detectors, are similar.
To test systematics, each correlation function for both polarization states and telescopes were fit by a Gaussian to extract the coherence time. 
Each individual fit was within 1$\,\sigma$ of the quoted coherence time, and furthermore the weighted mean of fits (1.35$\pm$0.05\,ps) is in perfect agreement with a single fit of co-added correlations indicating that within our measurement precision there are no significant systematics that preclude us from combining individual zero-baseline correlations from separate detector pairs.

The coherence time measured from the zero-baseline correlations can then be compared to expectation from the recorded spectrum. 
The spectral transmission of the H$\alpha$ filter was measured in the laboratory with a high resolution spectrograph~\citep{Matthews2022}.
Spectra of $\gamma$-Cas with resolution R=25000 were recorded contemporaneously with our observations using a {\sl Whoppshel} echelle spectrograph provided through collaboration with the 2SPOT\footnote{www.2spot.org} association of amateur astronomers. 
This is especially important as the width of the temporally variable emission line is narrower than the filter bandpass thus affecting the coherence time. 
The right side of Figure~\ref{fig:g2_r_0} shows the filter transmission, and the emission line spectra. Through Equation~\ref{eqn:coh_time} we extract the expected coherence time to be 1.41\,ps which is 1.2\,$\sigma$ larger than the value from zero-baseline correlations when taking only the measurement uncertainty and thus in fair agreement.
In contrast, the coherence time that would be expected for a flat stellar spectrum would be 1.16\,ps. 
This is considerably less than the measured value by 3.8$\,\sigma$ illustrating the importance of including the emission line profile in calculations of the coherence time.
The general agreement of the coherence time measured between intensity interferometry and spectral measurements indicate that there are no systematic effects arising from the presence of unidentified narrow spectral lines due to a lack of spectral resolution.

\subsection{Spatial Intensity Correlations}

The spatial intensity correlations correspond to the correlations between all detector pairs on separate telescopes that are observing in the same polarization mode corresponding to a total of 8 cross-correlations across telescopes. 
All computed correlations are shifted in time by instrumental and geometrical delays, and then co-added together. 
The full data set corresponds to a wide projected baseline and position angle range and so the data was divided into smaller baseline ranges. 
For each sub-division, we compute the averaged correlation function and then fit a Gaussian function to the resulting bunching peak. 
The measured areas of the $g^{(2)}$ peak for each subdivision are presented in Table~\ref{tab:gamcasobsparams}.
Squared visibilities are extracted by computing the ratio of the integral of the Gaussian peak of the cross-correlation to the computed value from the measured spectrum.

\begin{deluxetable}{cc}
\tablecaption{Spatial intensity correlation results. \label{tab:gamcasobsparams}
}
\tablehead{
\colhead{Baseline Range} & \colhead{Peak Area} \\
\colhead{m} & \colhead{ps}
}
\startdata
0.0 & 1.35$\pm$0.05 \\
3.8 $<$ r $\leq$ 13.0 & 1.40$\pm$0.17 \\
13.0 $<$ r $\leq$ 18.0 & 0.83$\pm$0.14 \\
18.0 $<$ r $\leq$ 21.3 & 0.31$\pm$0.08 \\
32.0 $<$ r $\leq$ 37.8 & 0.07$\pm$0.04 \\
\enddata
\end{deluxetable}

\section{Analysis of Results}
\label{sec:analysis}

The reduced II data resulted in 4 measurements of the squared visibility, each averaged over a range of baselines required to significantly resolve a bunching peak.
In turn, the limited sampling does not allow any reasonable independent visibility modeling. 
Nevertheless, it is interesting to compare the measured values to past results.
Past interferometric observations generally characterize the angular brightness distribution of $\gamma$-Cas with a parameterized geometrical model.
A common assumption is a two-component system consisting of a photosphere and disk, with some flux ratio between them. 
The photosphere is typically approximated as a uniform disk. 
This is an oversimplified model of Be stars as it does not take into strong temperature gradients and equatorial
flattening from the near critical rotation~\citep{DomicianodeSouza2002}.
However, to resolve these effects requires an angular resolution at the characteristic diameter of the photosphere ($\sim$ 0.5\,mas for $\gamma$-Cas \citep{Stee1998}), whereas the effective resolution for our observations (1.22$\lambda / D$) at the largest baseline is $\sim 8.3\,$mas.
Furthermore, these observations were conducted within the H$\alpha$ line in which the disk emission is much stronger, such that photospheric contributions are significantly minimized.
For the disk emission, several geometric models were tested, including elongated uniform disks, Gaussian disks, and uniform rings. \citet{Tycner2006,Stee2012} showed that a Gaussian disk profile best described the extended emission relative to the other assumptions.

\begin{deluxetable}{ccccc}[t]
\tablecaption{Reported Gaussian disk fit values in prior $\gamma$-Cas observations. $\theta_{GD}$ is the full-width at half-maximum, $\phi$ is the position angle, and $r$ is the axial ratio.  \label{tab:gamcas_params}
}
\tablehead{
\colhead{Observatory} &
\colhead{Ref.} & \colhead{$\theta_{GD}$} & \colhead{$\phi$} & \colhead{r} \\
\colhead{} &
\colhead{} & \colhead{(mas)} & \colhead{($^{\circ}$)} & \colhead{}
}
\startdata
MkIII & \cite{Quirrenbach1997} & 3.47$\pm$0.02 & 19$\pm$2 & 0.70$\pm$0.02 \\
NPOI & \cite{Tycner2006} & 3.59$\pm$0.04 & 31.2$\pm$1.2 & 0.58$\pm$0.03 \\
CHARA & \cite{Stee2012} & 4.4$\pm$0.4 & 19$\pm$5 & 0.74 \\
\enddata
\end{deluxetable}

\begin{figure}[t]
    \plottwo{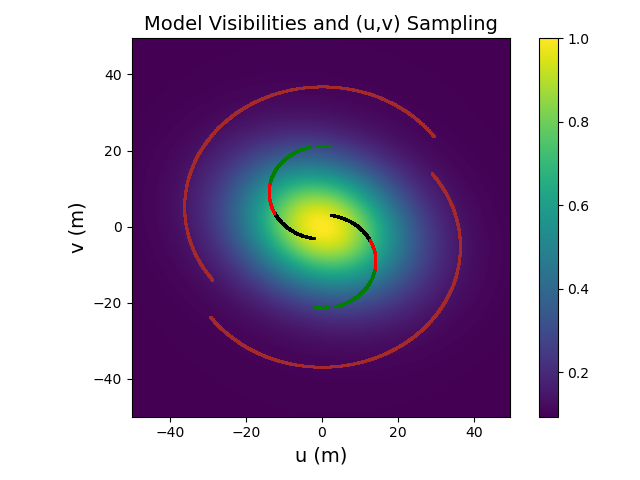}{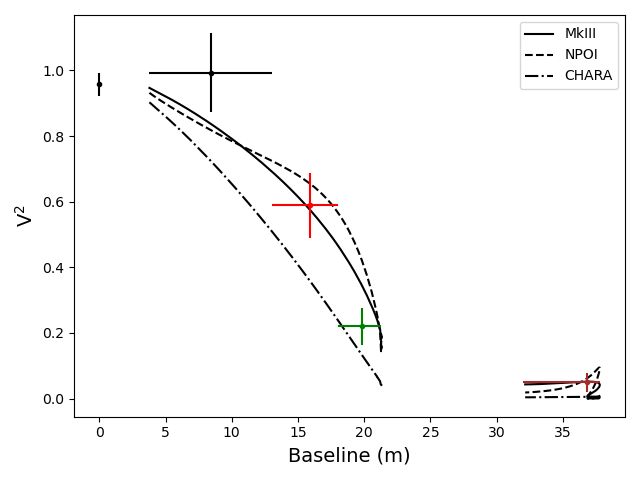}
    \caption{The left image shows the uv-plane coverage, over-plotted on expected squared visibilities formed from a Gaussian disk model of $\gamma$-Cas using reported parameters from \cite{Stee2012}. Each of the colors represent the range of sampled points averaged together in order to measure squared visibilities, as correspondingly plotted on the right.
    Additionally, we plot the expected squared visibilities for each of the models in Table~\ref{tab:gamcas_params} at our sampled uv-plane points.}
    \label{fig:gam_cas_visbs}
\end{figure}

This two component model of a uniform disk + elongated Gaussian disk was applied to $\gamma$-Cas data in three prior reported observations using the MkIII interferometer \citep{Quirrenbach1997}, NPOI \citep{Tycner2006}, and CHARA \citep{Stee2012}. 
The reported parameters for these observations are summarized and presented in Table~\ref{tab:gamcas_params}. 
Figure~\ref{fig:gam_cas_visbs} shows our squared visibilities, along with expected values obtained from a Gaussian disk model from the previous observations.
The comparison of our results with the models produced using reported values from literature tends to align with the values given by \cite{Tycner2006} and \cite{Quirrenbach1997} over \cite{Stee2012} who suggest a smaller extent of the H$\alpha$ region.
Within our measurement precision this is not strongly conclusive and can also be a result of instrumental differences. 
Already, \cite{Stee2012} noted the larger angular extent could be explained in that they used high resolution spectro-interferometry, in contrast to the other observations including our own, that utilize narrowband filters.
The reasoning is that the filters detect more of the less resolved continuum resulting in an effectively smaller angular extent. 

\section{Discussion and Outlook}
\label{sec:discussion}

We reported here on II measurements of the extended H$\alpha$ emitting region of $\gamma$-Cas. 
The observed angular extent of the emission was found to be consistent with past direct interferometry measurements.
Following our previous observations of \cite{Rivet2020} and \cite{deAlmeida2022} this extends the work of II measurements in emission lines to another system and complements recent on-sky results of other intensity interferometry facilities \citep{Acciari2020,Abeysekara2020,Horch2022}.
Future improvements to the system will aim to improve the sensitivity.
The most significant gain comes from simultaneously performing II correlations in many spectral channels that can be co-added to improve the signal to noise ratio by a factor of the square root of the number of channels \citep{Trippe2014}.
One could also imagine recording many independent spectral channels across the H$\alpha$ line in order to perform intensity spectro-interferometry to test the discrepancy seen in past observations between those using filtered bandpasses, and those in dispersed light.
Furthermore, higher sensitivity observations, paired with polarimetric capabilities would allow for better constraints, if not a direct measure of, radiative processes displaying polarized emission in the disk, as was attempted by \cite{RousseletPerraut1997}.

These observations were performed using two facilities: the T1M and M\'{e}O that had not been used for interferometric observations prior to this report. The portability of the T1M provides the capability to optimize the baseline configuration for a given target, or similarly perform multiple configurations for a given target as done here. This technical accomplishment also illustrates the potential for performing up to 4-telescope II measurements on the Calern Plateau enabling 6 simultaneous baselines by including the 2 additional C2PU telescopes. 

\acknowledgments

We acknowledge the financial support of the Région PACA (project I2C), the French National Research Agency (ANR, project I2C, ANR-20-CE31-0003), OCA, Doeblin federation, UCA science councils grants, and the LABEX Cluster of Excellence FIRST-TF (ANR-10-LABX-48-01), within the program Investissements d'Avenir operated by the ANR.
The authors would like to thank Jacques Belin and Damien Pesce for their installment of reference markers, and to the members of the M\`{e}O team including Hervey Mariey, Mourad Aimar, Herv\'{e} Viot, Gr\'{e}goire Martinot-Lagarde, Julien Scariot, Nicolas Maurice, Duy-H\`{a} Phung, and Nils Raymond for their assistance during the observations.

\end{document}